\documentclass[aps,pra,twocolumn]{revtex4}%
\usepackage{graphics}
\usepackage{amsmath}
\usepackage{graphicx}
\usepackage{amsfonts}
\usepackage{amssymb}
\usepackage{float}
\usepackage{longtable}
\usepackage{epsfig}
\usepackage{latexsym}
\usepackage{theorem}
\usepackage{tikz}
\usepackage{bbm}
\usepackage{bm}
\usepackage{braket}
\usepackage{psfrag}

\def\^PT{^{\mathrm{PT}}}
\begin{document}
\title{Adaptive estimation and discrimination of Holevo-Werner channels}
\author{Thomas P. W. Cope}
\author{Stefano Pirandola}
\affiliation{Computer Science and York Centre for Quantum Technologies, University of York,
York YO10 5GH, UK}

\begin{abstract}
The class of quantum states known as Werner states have several interesting
properties, which often serve to illuminate unusual properties of quantum
information. Closely related to these states are the Holevo-Werner channels
whose Choi matrices are Werner states. Exploiting the fact that these channels
are teleportation covariant, and therefore simulable by teleportation, we
compute the ultimate precision in the adaptive estimation of their
channel-defining parameter. Similarly, we bound the minimum error probability
affecting the adaptive discrimination of any two of these channels. In this
case, we prove an analytical formula for the quantum Chernoff bound which also
has a direct counterpart for the class of depolarizing channels. Our work
exploits previous methods established in [Pirandola and Lupo, PRL
\textbf{118}, 100502 (2017)] to set the metrological limits associated with
this interesting class of quantum channels at any finite dimension.

\end{abstract}
\maketitle

\section{Introduction}

When asked about the advances quantum information
technology~\cite{NiCh,QIbook,RMP,BraRMP,Ulrikreview,Kimble,HybridINTERNET}
will make in the future, most commonly mentioned will be quantum
cryptography~\cite{crypt1,crypt2,crypt3,crypt5,diamantiENTROPY,usenkoREVIEW,Grosshans2003b,
weedbrook2004noswitching,pirs2way,MDI1,MDILo,CVMDIQKD} or the potential
advances of quantum computing~\cite{comp1,comp4,comp5,NiCh}. Despite this, one
of the fastest growing areas is that of \textit{quantum metrology}%
~\cite{met1,met2,met3,met4,met5,met6,Paris,Makei,Lupo16,Nair,Metro}, where
parameters of physical systems are estimated with high precision, often using
resources such as entangled or spin-squeezed states to achieve higher
resolution. The two bounds often stated in metrology are the \textit{standard
quantum limit}, in which the error variance associated with the parameter
estimation scales as $n^{-1}$, with $n$ being the number of uses, and the
\textit{Heisenberg limit}, with improved scaling of $n^{-2}$.

Another important area is that of quantum hypothesis
testing~\cite{QHT,QHT2,Invernizzi,QHB1,Gae1} and its formulation in terms of
quantum channel discrimination. The latter is particularly important in
problems of quantum sensing, e.g., in quantum
reading~\cite{Qread,Qread2,Qread3,Qread4,Nair11,Hirota11,Bisio11,Arno11} or in
quantum illumination~\cite{Qill0,Qill1,Qill2,Qill3,Qill4,Qill5}. When the
discrimination problem is binary (i.e., with two hypotheses) and symmetric
(i.e., with the same Bayesian costs), the main tool is the Helstrom
bound~\cite{Hesltrom} which reduces the computation of the minimum error
probability to the trace distance~\cite{NiCh}. Notable lower and upper bounds
to the probability can also be expressed in terms of the
fidelity~\cite{Fid1,Fid2,Banchi} and the quantum Chernoff
bound~\cite{QCB1,QCB2,QCB3}, which are particularly useful when many copies
are considered in the discrimination process.

Recently, Ref.~\cite{Metro} showed how quantum
teleportation~\cite{tele,Samtele,teleREVIEW} is a primitive
operation in the fields of quantum metrology and quantum
hypothesis testing. First of all, whenever a quantum channel is
teleportation-covariant~\cite{Stretching}, i.e., it suitably
commutes with the random unitaries of teleportation, it can be
simulated by teleporting over its Choi matrix (see
Ref.~\cite{RevewTQC} for a review). As shown in Ref.~\cite{Metro},
this channel simulation can then be exploited to re-organise the
most general possible \textit{adaptive} protocol of channel
estimation/discrimination into a much simpler block version, where
the unknown channel is probed in an independent and identical
fashion up to some general quantum operation. Thanks to this
reduction, one can compute the ultimate limit in the adaptive
estimation or discrimination of noise parameters encoded in
teleportation-covariant channels. This family includes Pauli
channels (depolarizing, dephasing), erasure channels, and also
bosonic Gaussian channels~\cite{Metro}.

In this manuscript, we adopt this recent methodology to study the
ultimate metrological limits of another class of
teleportation-covariant channels:\ the Holevo-Werner (HW)
channels, defined as those channels whose Choi matrices are Werner
states~\cite{Werner,Barrett,NPTstates}. They hold an important
place in quantum information, since one element of this class was
used to disprove the conjecture of the additivity of minimal
Reny\'{\i} entropy~\cite{counter}. As with the class of Werner
states, the HW channels can be parametrised by a real parameter
$\eta\in\lbrack-1,1]$, and we use the notation $\mathcal{W}_{\eta
,d}:\mathcal{H}_{d}\rightarrow\mathcal{H}_{d}$ at dimension $d$.

By using the quantum Fisher information (QFI) and the quantum Cramer-Rao bound
(QCRB)~\cite{met1,CRbound}, we then compute the ultimate precision in the
adaptive estimation of the channel-defining parameter $\eta$. The analytical
formula is simple and the bound is asymptotically achievable by a non-adaptive
strategy. Then, we consider the adaptive discrimination of two
(iso-dimensional) HW channels with arbitrary parameters $\eta$ and $\zeta$.
The minimum error probability can be bounded by single-letter quantities in
terms of the fidelity, the relative entropy and the quantum Chernoff bound
(QCB)~\cite{QCB1}. For the latter, we show an analytical formula and a
corresponding one for the class of standard depolarizing channels.

The structure of this paper is as follows. In Sec.~\ref{Wer_preli}, we
describe Werner states and HW channels, also explaining their teleportation
covariance. In Secs.~\ref{MetroSec} and~\ref{DISCsection} we then derive the
ultimate metrological and discrimination limits associated with these
channels, giving explicit analytical formulas. We then conclude in
Sec.~\ref{Conclu}.

\section{Holevo-Werner channels and their properties\label{Wer_preli}}

Werner states~\cite{Werner} are defined over two qudits of equal dimension
$d$. They have the peculiar property that they are invariant under local
unitaries
\begin{equation}
(U_{d}\otimes U_{d})W_{\eta,d}(U_{d}^{\dagger}\otimes U_{d}^{\dagger}%
)=W_{\eta,d}. \label{invar}%
\end{equation}
Whilst there exist several parametrisations of this family, here
we shall use the expectation representation, so that
\begin{equation}
\eta:=\mathrm{Tr}\left(  W_{\eta,d}\mathbb{F}\right)  , \label{Tracecon}%
\end{equation}
where $\mathbb{F}$ is the flip operator acting on two qudits, i.e.,
\begin{equation}
\mathbb{F}:=\sum_{i,j=0}^{d-1}\ket{ij}\bra{ji}~,
\end{equation}
with $\{\ket{i}\}$ being the computational basis. The expectation $\eta$
ranges from $-1$ to $1$, with separable Werner states having nonnegative expectations.

We also have an explicit formula for $W_{\eta,d}$ as a linear combination of
the $\mathbb{F}$ operator and the $d^{2}\times d^{2}$ identity operator
$\mathbb{I}$, i.e.,~\cite{Werner}
\begin{equation}
W_{\eta,d}=\frac{(d-\eta)\mathbb{I}+(d\eta-1)\mathbb{F}}{d^{3}-d},
\label{stateform}%
\end{equation}
from which Eq.~(\ref{Tracecon}) is easy to verify. It is known
that, for $d\geq3$, there exist Werner states which are entangled
but yet admit a local model for all
measurements~\cite{Werner,Barrett}. Also, the extremal entangled
Werner state $W_{-1,d}$ was used to disprove~\cite{VolWer} the
additivity of the relative entropy of entanglement
(REE)~\cite{VedFORM,Plenio,VedralRMP}. Werner states of a given
dimension $d$ have a nice property: for any value $\eta$, they
share the same eigenbasis, so that they are simultaneously
diagonalisable. In particular, a Werner state $W_{\eta,d}$ has the
following eigenspectrum: $d(d+1)/2$ eigenvectors with eigenvalue
$(1+\eta)[d(d+1)]^{-1}$ and $d(d-1)/2$ eigenvectors with
eigenvalue $(1-\eta)[d(d-1)]^{-1}$.

Recall that the Choi matrix of a quantum channel $\mathcal{E}:\mathcal{H}%
_{d}\rightarrow\mathcal{H}_{d}$ is defined as $\chi_{\mathcal{E}}%
:=\mathbf{I}\otimes\mathcal{E}(\ket{\Phi}\bra{\Phi})$, where
$\ket{\Phi}=d^{-1/2}\sum_{i=0}^{d-1}\ket{ii}$ is a maximally-entangled state
and $\mathbf{I}$\ is the $d$ dimensional identity map. Then, the HW\ channels
are those channels whose Choi matrices are the Werner states, i.e.,
$\chi_{\mathcal{W}_{\eta,d}}=W_{\eta,d}$. Their action on an input state
$\rho$ is given by~\cite{Fanneschannel}
\begin{equation}
\mathcal{W}_{\eta,d}\left(  \rho\right)  :=\frac{(d-\eta)\mathbf{I}%
+(d\eta-1)\rho^{T}}{d^{2}-1}, \label{channeldef}%
\end{equation}
with $\rho^{T}$ the transposed state. In particular, the extremal HW channel
\begin{equation}
\mathcal{W}_{-1,d}\left(  \rho\right)  =\frac{\mathbf{I}-\rho^{T}}{d-1}%
\end{equation}
is one-to-one with the extremal Werner state $W_{-1,d}$. The latter channel
was used as a counterexample of the additivity of minimal Reny\'{\i} entropy
\cite{counter} whilst the minimal output entropy of (\ref{channeldef}) was
proven to be additive~\cite{Fanneschannel}.

For completeness, recall that closely related to Werner states are the
isotropic states~\cite{horoISO}, defined by
\begin{equation}
\Omega_{\alpha,d}=\frac{(d-\alpha)\mathbb{I}+(d\alpha-1)\mathbb{M}}{d^{3}-d}
\label{ptstate}%
\end{equation}
where $\mathbb{M}$ is the maximally entangled operator
\begin{equation}
\mathbb{M}:=\sum_{i,j=0}^{d-1}\ket{ii}\bra{jj},
\end{equation}
and $\alpha:=\mathrm{Tr}(\Omega_{\alpha,d}\mathbb{M})$\ ranges in $\left[
0,d\right]  $. For $\alpha\leq1$, we have a separable isotropic state. The
latter can be formed by taking the partial transpose (PT) of a (separable)
Werner state with $\eta=\alpha$, i.e.,
\begin{equation}
\Omega_{\alpha,d}=W_{\alpha,d}^{\mathrm{PT}}~~(\alpha\leq1).
\end{equation}
Isotropic states of a given dimension are also simultaneously
diagonalisable. Their eigenspectrum has $1$ eigenvector with
eigenvalue $\eta/d$ and $d^{2}-1$ eigenvectors with eigenvalue
$(d-\eta)[d(d^{2}-1)]^{-1}$.

It is known that the isotropic state $\Omega_{\alpha,d}$ is the Choi matrix of
a depolarizing channel $\mathcal{D}_{\alpha,d}$, whose action is~\cite{NiCh}
\begin{equation}
\mathcal{D}_{\alpha,d}\left(  \rho\right)  :=\frac{(d-\alpha)\mathbf{I}%
+(d\alpha-1)\rho}{d^{2}-1}~. \label{channeldefisoB}%
\end{equation}
Representing the isotropic state as $\Omega_{p,d}=pd^{-2}\mathbb{I}%
+(1-p)\ket{\Phi}\bra{\Phi}$ with $p\in\left[  0,d^{2}/(d^{2}-1)\right]  $,
then we may write $\mathcal{D}_{p,d}\left(  \rho\right)  =p\frac{\mathbf{I}%
}{d}+(1-p)\rho$. In fact, we may easily convert between the two forms by using
$p=d(d-\alpha)(d^{2}-1)^{-1}$. From Eqs.~(\ref{channeldef})
and~(\ref{channeldefisoB}), we see that depolarizing and HW channels are
equivalent up to a transposition, which is why we may also call the HW
channels as \textquotedblleft transpose\textquotedblright\ depolarizing channels.

Like depolarizing channels, HW channels are also teleportation-covariant.
Recall that a quantum channel $\mathcal{E}$ is called \textquotedblleft
teleportation covariant\textquotedblright\ if, for every teleportation unitary
$U$ (i.e., Pauli unitary in finite dimension~\cite{tele} and displacement
operator in infinite dimension~\cite{teleREVIEW,Samtele}), there exists some
unitary $V$ such that
\begin{equation}
\mathcal{E}\left(  U\rho U^{\dagger}\right)  =V\mathcal{E}\left(  \rho\right)
V^{\dagger}. \label{telecovv}%
\end{equation}
This concept was discussed in Refs.~\cite{MHthesis,Wolfnotes,Leung} for
discrete variable systems, and generally formulated in Ref.~\cite{Stretching}
for both the discrete and continuous variables (see also Ref. \cite{RevewTQC}
for a review). One can check that the HW channels are teleportation-covariant.
For an arbitrary unitary $U$, we have $\mathbf{I}=U^{\ast}\mathbf{I}(U^{\ast
})^{\dagger}$ and $(U\rho U^{\dagger})^{T}=U^{\ast}\rho^{T}(U^{\ast}%
)^{\dagger}$. Therefore, from Eq.~(\ref{channeldef}), we find that
\begin{equation}
\mathcal{W}_{\eta,d}\left(  U\rho U^{\dagger}\right)  =U^{\ast}\mathcal{W}%
_{\eta,d}\left(  \rho\right)  (U^{\ast})^{\dagger},
\end{equation}
which realises Eq.~(\ref{telecovv}) with $V=U^{\ast}$.

Because HW channels are teleportation covariant, they can be simulated by
teleporting over their Choi matrices~\cite{Stretching,RevewTQC}. Let us call
$\mathcal{T}_{d}$ the local operations and classical communication (LOCC)
associated with the $d$-dimensional teleportation protocol. Then, we may write%
\begin{equation}
\mathcal{W}_{\eta,d}\left(  \rho\right)  =\mathcal{T}_{d}\left(  \rho\otimes
W_{\eta,d}\right)  .\label{channelSIMK}%
\end{equation}
More precisely, the HW channel $\mathcal{W}_{\eta,d}$ forms a class of jointly
teleportation-covariant channels with respect to the parameter $\eta$. This
means that $\mathcal{W}_{\eta,d}$ satisfies Eq.~(\ref{telecovv}) with the
output unitaries $V$ independent of $\eta$. For this reason, in the channel
simulation in Eq.~(\ref{channelSIMK}), the parameter $\eta$ only appears as a
noise parameter in the Choi matrix and not in the teleportation LOCC
$\mathcal{T}_{d}$.

Using the simulation in Eq.~(\ref{channelSIMK}), an adaptive protocol over $n$
uses of the HW channel $\mathcal{W}_{\eta,d}$ can be reduced to a block
protocol over a tensor product of Werner states $W_{\eta,d}^{\otimes n}$. In
the literature~\cite{RevewTQC}, this type of adaptive-to-block simplification
was first introduced for the tasks of quantum/private communications in
Ref.~\cite{Stretching}. See also
Refs.~\cite{ref1,multipoint,RicFINITE,Tompaper}. Later it was extended to
quantum metrology and channel discrimination~\cite{Metro}. See
Ref.~\cite{ReviewMETRO} for a review on channel simulation and adaptive metrology.

\section{Quantum parameter estimation with Holevo-Werner
channels\label{MetroSec}}

Consider a HW channel $\mathcal{W}_{\eta,d}$\ with known dimension
$d$ but unknown parameter $\eta$. The most general parameter
estimation protocol is adaptive and consists of $n$ probings of
the channel, interleaved by quantum operations~\cite{Metro}. In
fact, we may assume that we use a register of quantum systems,
from which we extract a system for each transmission through the
channel. After each transmission, the output is re-combined with
the register which is then subject to a global quantum operation.
This is repeated $n$ times, after which the state of the register
$\rho_{\eta}^{n}$ is measured, and the outcome is processed into
an optimal unbiased estimator $\tilde{\eta}$ of $\eta$. The
minimum error probability $\mathrm{Var}\left( \eta\right)
:=\left\langle (\eta-\tilde{\eta})^{2}\right\rangle $ satisfies
the QCRB~\cite{met1}%
\begin{equation}
\mathrm{Var}\left(  \eta\right)  \geq(\bar{I}_{\eta}^{n})^{-1},
\end{equation}
where $\bar{I}_{\eta}^{n}$ is the QFI optimised over all the
adaptive protocols $\mathcal{P}$. More precisely, this
optimisation is over all possible input states and quantum
operations for the register, and over all possible output
measurements. In terms of the Bures' quantum fidelity
$F(\rho,\sigma):=\mathrm{Tr}\sqrt{\sqrt{\sigma}\rho\sqrt{\sigma}}$,
we may
write the following expression~\cite{Metro}%
\begin{equation}
\bar{I}_{\eta}^{n}:=\sup_{\mathcal{P}}\frac{8\left[  1-F(\rho_{\eta}^{n}%
,\rho_{\eta+\delta\eta}^{n})\right]  }{\delta\eta^{2}}~, \label{defSTART}%
\end{equation}
where $\rho_{\eta}^{n}$ is the output of protocol $\mathcal{P}$.

Because the HW channel $\mathcal{W}_{\eta,d}$ is (jointly) teleportation
covariant and therefore simulable by teleporting over its Choi matrix
$W_{\eta,d}$ (which is a Werner state) with a $\eta$%
-independent\ teleportation LOCC $\mathcal{T}_{d}$ as in
Eq.~(\ref{channelSIMK}), we may re-organise any adaptive protocol
of parameter estimation into a block protocol so that the output
state of the register takes the form~\cite{Metro}
\begin{equation}
\rho_{\eta}^{n}=\bar{\Lambda}(W_{\eta,d}^{\otimes n})\label{newEQ}%
\end{equation}
for a trace-preserving quantum operation $\bar{\Lambda}$ not
depending on the parameter $\eta$ (see~\cite{ReviewMETRO} for more
details on how adaptive protocols of quantum metrology may be
fully simplified). This allows us to simplify the QFI, which
becomes a function of the Choi matrix $W_{\eta,d}$.
Following~\cite{Metro}, we may remove the supremum in
Eq.~(\ref{defSTART}) and simplify the formula to the following
\begin{equation}
\bar{I}_{\eta}^{n}=8n\frac{1-F(W_{\eta,d},W_{\eta+\delta\eta,d})}{\delta
\eta^{2}}.\label{QFIformula}%
\end{equation}

For the sake of clarity let us briefly repeat the steps of the proof of
Ref.~\cite{Metro} for our specific case. Eq.~(\ref{QFIformula}) can be proven
by combining Eq.~(\ref{newEQ}) with basic properties of the fidelity, i.e.,
(i)~monotonicity under $\bar{\Lambda}$ and (ii)~multiplicativity over tensor
products. In fact, we may write%
\begin{align}
&  F(\rho_{\eta}^{n},\rho_{\eta+\delta\eta}^{n})\overset{\text{(i)}}{\geq
}F(W_{\eta,d}^{\otimes n},W_{\eta+\delta\eta,d}^{\otimes n})\nonumber\\
&  \overset{\text{(ii)}}{=}F(W_{\eta,d},W_{\eta+\delta\eta,d})^{n}.
\end{align}
Note that all the information about the protocol $\mathcal{P}$ was contained
in $\bar{\Lambda}$, which disappears in the inequality above. We have
therefore the upper bound%
\begin{equation}
\bar{I}_{\eta}^{n}\leq B(n):=\frac{8(1-F^{n})}{\delta\eta^{2}},~F:=F(W_{\eta
,d},W_{\eta+\delta\eta,d}).
\end{equation}

As in Ref.~\cite{Metro}, we now show that the upper bound $B(n)$ is additive.
For $n=1$ and $\delta\eta\rightarrow0$, we have $F=1-B(1)\delta\eta^{2}/8$
implying $F^{n}=1-nB(1)\delta\eta^{2}/8+O(\delta\eta^{4})$. Up to higher order
terms, the latter expansion implies the additivity $B(n)=nB(1)$, so that we
may write
\begin{equation}
\bar{I}_{\eta}^{n}\leq nB(1)=8n\frac{(1-F)}{\delta\eta^{2}}.\label{kkkl}%
\end{equation}
The next step is to show the achievability of the upper bound in the latter
inequality. Consider a block protocol $\mathcal{\tilde{P}}$ where we prepare
$n$ maximally-entangled states $\Phi^{\otimes n}=\ket{\Phi}\bra{\Phi}^{\otimes
n}$ and partly propagate them through the channel, so that the output is equal
to $\rho_{\eta}^{n}=W_{\eta,d}^{\otimes n}$. It is easy to see that this
specific protocol achieves the QFI $I_{\eta}^{n}(\mathcal{\tilde{P}})=nB(1)$,
thus $\bar{I}_{\eta}^{n}=nB(1)$, completing the proof of Eq.~(\ref{QFIformula}%
). Also note that because $\mathcal{\tilde{P}}$ uses independent input states,
the QCRB $\mathrm{Var}(\eta)\geq\lbrack I_{\eta}^{n}(\mathcal{\tilde{P}%
})]^{-1}$ is achievable for large $n$ via local measurements~\cite{Gillapp}.

Thus, the problem is reduced to computing the fidelity between two
Werner states. Because these states are diagonalisable in the same
basis, we may write
\begin{equation}
F(W_{\eta,d},W_{\zeta,d})=\sum_{i}\sqrt{p_{i}q_{i}},
\end{equation}
where $p_{i}$ and $q_{i}$ are the eigenvalues of $W_{\eta,d}$ and $W_{\zeta
,d}$ respectively. After some algebra, we find
\begin{equation}
F(W_{\eta,d},W_{\zeta,d})=\frac{\sqrt{(1+\eta)(1+\zeta)}}{2}+\frac
{\sqrt{(1-\eta)(1-\zeta)}}{2}.\label{fidelity}%
\end{equation}

Now it is easy to see that the Taylor expansion of $1-F\left[  W_{\eta
,d},W_{\eta+\delta\eta,d}\right]  $ around $\delta\eta\approx0$ provides
\begin{equation}
\frac{\left(  1-F\left[  W_{\eta,d},W_{\eta+\delta\eta,d}\right]  \right)
}{\delta\eta^{2}}=\frac{1}{8(1-\eta^{2})}.
\end{equation}
Substituting this into Eq.~(\ref{QFIformula}), we derive%
\begin{equation}
\bar{I}_{\eta}^{n}=\frac{n}{1-\eta^{2}},
\end{equation}
so that the QCRB is given by
\begin{equation}
\mathrm{Var}\left(  \eta\right)  \geq\frac{1-\eta^{2}}{n}~.
\end{equation}

Here we may make several observations. First of all, we notice that the QCRB
is surprisingly dimension-independent. Second, as expected from teleportation
covariant channels, we cannot beat the standard quantum limit. Third, this
bound is also asymptotically achievable for large $n$. In fact, as already
said in the previous proof, a specific strategy consists in probing the
channel (identically and independently) with part of maximally-entangled states.

\section{Bounds for adaptive channel discrimination\label{DISCsection}}

Consider now the problem of symmetric binary discrimination with two
equiprobable (and iso-dimensional) HW channels $\mathcal{E}_{0}=\mathcal{W}%
_{\eta,d}$ and $\mathcal{E}_{1}=\mathcal{W}_{\zeta,d}$. The
unknown channel $\mathcal{E}_{u}$ (with $u=0,1$) is stored in a
box which is probed $n$ times according to an adaptive
discrimination protocol~\cite{Metro}. This protocol is as the one
described before for parameter estimation but tailored for the
different task of discrimination. In particular, this means that
the output state $\rho_{u}^{n}$ encodes the bit of information $u$
associated with the two hypotheses, and is subject to a dichotomic
Helstrom measurement~\cite{Hesltrom}. The mean error probability
affecting the discrimination is therefore expressed in terms of
the Helstrom
bound~\cite{Hesltrom}, i.e., $p_{\text{err}}=[1-D(\rho_{0}^{n},\rho_{1}%
^{n})]/2$ where $D$ is the trace distance~\cite{NiCh}. By
minimising over all
adaptive protocols, we define the optimal error probability $p_{\text{err}%
}^{\text{opt}}$.

Because the two iso-dimensional HW channels $\mathcal{W}_{\eta,d}$
and $\mathcal{W}_{\zeta,d}$ are jointly teleportation covariant,
i.e., we may write Eq.~(\ref{telecovv}) with exactly the same set
of output unitaries $V$, then the two channels are
teleportation-simulable with exactly the same teleportation LOCC
$\mathcal{T}_{d}$ (but over different Choi matrices $W_{u,d}$).
For this reason, we may re-organise the adaptive discrimination
protocol into a block protocol with output state
$\rho_{u}^{n}=\bar{\Lambda }(W_{u,d}^{\otimes n})$ for a
($u$-independent) trace-preserving quantum operation
$\bar{\Lambda}$. This allows us to write single-letter bounds for
$p_{\text{err}}^{\text{opt}}$. In particular, we have~\cite{Metro}
\begin{equation}
\frac{1-\sqrt{\min\{1-F^{2n},nS\}}}{2}\leq p_{\text{err}}^{\text{opt}}%
\leq\frac{Q^{n}}{2}\leq\frac{F^{n}}{2}, \label{discBB}%
\end{equation}
where $F:=F(W_{\eta,d},W_{\zeta,d})$, $Q$ is the quantum Chernoff bound (QCB)
\begin{equation}
Q:=\inf_{s\in\lbrack0,1]}\mathrm{Tr}\left[  W_{\eta,d}^{s},W_{\zeta,d}%
^{1-s}\right]  , \label{FullQCB}%
\end{equation}
and $S$ is related to the relative entropy
\begin{equation}
S:=(\ln\sqrt{2})\min\{S\left(  W_{\eta,d}||W_{\zeta,d}\right)  ,S\left(
W_{\zeta,d}||W_{\eta,d}\right)  \}. \label{Sdef}%
\end{equation}

Whilst these bounds may seem complicated, we have analytical formulae for each
of these quantities. We have already seen the fidelity in Eq.~(\ref{fidelity})
between two Werner states, which can be used here for the fidelity bounds in
Eq.~(\ref{discBB}). Then, we may also compute
\begin{equation}
S=%
\begin{cases}
(\mathrm{ln}\sqrt{2})\left(  \frac{1+\eta}{2}\mathrm{log}_{2}\frac{1+\eta
}{1+\zeta}+\frac{1-\eta}{2}\mathrm{log}_{2}\frac{1-\eta}{1-\zeta}\right)  &
~|\eta|\geq|\zeta|,\\
(\mathrm{ln}\sqrt{2})\left(  \frac{1+\zeta}{2}\mathrm{log}_{2}\frac{1+\zeta
}{1+\eta}+\frac{1-\zeta}{2}\mathrm{log}_{2}\frac{1-\zeta}{1-\eta}\right)  &
~|\eta|\leq|\zeta|.
\end{cases}
\label{SSs}%
\end{equation}

To find this, we first calculate the relative entropy between two
Werner states. Diagonalising them in the same basis, we may write
\begin{align}
S(W_{\eta,d}||W_{\zeta,d})  &  :=\mathrm{Tr}(W_{\eta,d}\mathrm{log}_{2}%
W_{\eta,d}-W_{\eta,d}\mathrm{log}_{2}W_{\zeta,d})\nonumber\\
&  =\sum_{i}p_{i}\mathrm{log}{\frac{p_{i}}{q_{i}}},
\end{align}
where $p_{i}$ are the eigenvalues of $W_{\eta,d}$ and $q_{i}$ are those of
$W_{\zeta,d}$. This allows us to compute
\begin{equation}
S(W_{\eta,d}||W_{\zeta,d})=\frac{1+\eta}{2}\mathrm{log}_{2}\frac{1+\eta
}{1+\zeta}+\frac{1-\eta}{2}\mathrm{log}_{2}\frac{1-\eta}{1-\zeta}.
\label{WernRel}%
\end{equation}

Using Eq.~(\ref{WernRel}), we can then evaluate%
\begin{align}
&  \Delta S:=S(W_{\eta,d}||W_{\zeta,d})-S(W_{\zeta,d}||W_{\eta,d})\nonumber\\
=  &  \left(  1+\frac{\eta+\zeta}{2}\right)  \mathrm{log}_{2}\frac{1+\eta
}{1+\zeta}\nonumber\label{entropycomp}\\
&  +\left(  1-\frac{\eta+\zeta}{2}\right)  \mathrm{log}_{2}\frac{1-\eta
}{1-\zeta}.
\end{align}
We can see that $\Delta S=0$ when $\eta=\pm\zeta$. We can study $\Delta S$ for
the valid regions of $\eta$ and $\zeta$, and (numerically) check that $\Delta
S<0$ for $|\eta|>|\zeta|$. This implies Eq.~(\ref{SSs}).

\begin{figure*}[ptb]
\vspace{-1.5cm}
\par
\begin{center}
\includegraphics[width=0.95\textwidth]{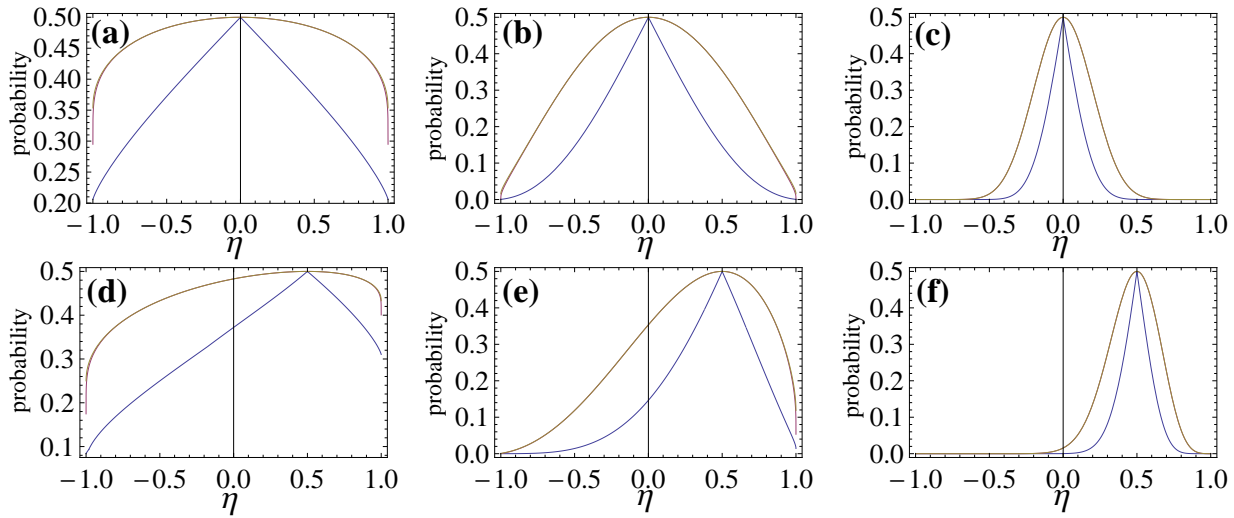}
\end{center}
\par
\vspace{-1.5cm}\caption{We plot the fidelity-based lower bound and the QCB
(upper bound) to the optimal error probability $p_{\text{err}}^{\text{opt}}$
in Eq.~(\ref{discBB}) for two HW channels $\mathcal{W}_{\eta,d}$ and
$\mathcal{W}_{\zeta,d}$ in arbitrary finite dimension $d\geq2$. In panels
(a)-(c), we set $\zeta=0$ and we plot the bounds as a function of $\eta$,
considering (a) $n=1$, (b) $n=10$, and (c) $n=100$. In panels (d)-(f), we
repeat the study with the same parameters as before but setting $\zeta=1/2$.}%
\label{error1}%
\end{figure*}

We now compute the QCB. The minimum error probability in the $n$-use\ adaptive
discrimination of two arbitrary HW channels $\mathcal{W}_{\eta,d}$ and
$\mathcal{W}_{\zeta,d}$\ is bounded by the QCB as in Eqs.~(\ref{discBB})
and~(\ref{FullQCB}), where $Q:=Q(W_{\eta,d},W_{\zeta,d})$ is computed as the
QCB between two corresponding Werner states $W_{\eta,d}$ and $W_{\zeta,d}$. We
find
\begin{align}
Q(W_{\eta,d},W_{\zeta,d})  &  =\inf_{s\in\lbrack0,1]}\left[  \frac{1+\zeta}%
{2}\left(  \frac{1+\eta}{1+\zeta}\right)  ^{s}\right. \nonumber\label{QCB}\\
&  \left.  +\frac{1-\zeta}{2}\left(  \frac{1-\eta}{1-\zeta}\right)
^{s}\right]  ,
\end{align}
where the infimum is analytically achieved at
\begin{equation}
s=%
\begin{cases}
\frac{1}{2}, & \eta=\zeta,\\
0^{+} & \eta=\pm1,\\
1^{-} & \zeta=\pm1,\\
\frac{\mathrm{ln}\left(  \frac{\zeta-1}{\zeta+1}\frac{\mathrm{ln}\frac{1-\eta
}{1-\zeta}}{\mathrm{ln}\frac{1+\eta}{1+\zeta}}\right)  }{\mathrm{ln}%
\frac{(1+\eta)(1-\zeta)}{(1+\zeta)(1-\eta)}}, & \text{otherwise.}%
\end{cases}
\label{mins}%
\end{equation}

In fact, since we may diagonalise two Werner states in the same
basis, Eq.~(\ref{FullQCB}) simplifies to
$\sum_{i}p_{i}^{s}q_{i}^{1-s}$, with $p_{i},q_{i}$ the eigenvalues
of $W_{\eta,d}$ and $W_{\zeta,d}$, respectively. We then minimise
this quantity by finding the unique turning point in $[0,1]$ and
showing it is indeed a minimum. The border points need a careful
consideration because they may show discontinuities and one needs
to take left or right limits. This is explicitly done in
Appendix~\ref{QCB_Werner}. In Fig.~\ref{error1}, we show numerical
examples on how the fidelity-based lower bound and the QCB [see
Eq.~(\ref{discBB})] behave in terms of $n$ for different values of
the channel-defining parameters $\eta$ and $\zeta$ for arbitrary
finite dimension $d\geq2$.

Using the same approach, we may also find an equivalent result for
depolarizing channels and their Choi matrices (isotropic states). The minimum
error probability in the $n$-use\ adaptive discrimination of two arbitrary
depolarizing channels $\mathcal{D}_{\alpha,d}$ and $\mathcal{D}_{\beta,d}$\ is
bounded by $p_{\text{err}}^{\text{opt}}\leq Q^{n}/2$ where $Q:=Q(\Omega
_{\alpha,d},\Omega_{\beta,d})$ is computed as the QCB between two
corresponding isotropic states $\Omega_{\alpha,d}$ and $\Omega_{\beta,d}$. We
find
\begin{equation}
Q(\Omega_{\alpha,d},\Omega_{\beta,d})=\inf_{s\in\lbrack0,1]}\left[
\frac{\beta}{d}\left(  \frac{\alpha}{\beta}\right)  ^{s}\right.  \left.
+\frac{d-\beta}{d}\left(  \frac{d-\alpha}{d-\beta}\right)  ^{s}\right]  ,
\end{equation}
where the infimum is analytically achieved at
\begin{equation}
s=%
\begin{cases}
\frac{1}{2} & \alpha=\beta,\\
0^{+} & \alpha=0,d,\\
1^{-} & \beta=0,d,\\
\frac{\mathrm{ln}\left(  \frac{\beta-d}{\beta}\frac{\mathrm{ln}\frac{d-\alpha
}{d-\beta}}{\mathrm{ln}\frac{\alpha}{\beta}}\right)  }{\mathrm{ln}\frac
{\alpha(d-\beta)}{\beta(d-\alpha)}} & \text{otherwise.}%
\end{cases}
\end{equation}
See Appendix~\ref{QCB_iso} for mathematical details. This bound for two
depolarizing channels is tighter than the fidelity-based upper bound of
Ref.~\cite{Metro}.

\section{Conclusion\label{Conclu}}

In this work, for the first time, we have considered Holevo-Werner
channels in the context of quantum metrology and quantum channel
discrimination, employing the most general (adaptive) protocols.
Because these channels are teleportation-covariant, the optimal
estimation of their channel-defining parameter $\eta$ is bounded
by the standard quantum limit, with an asymptotically achievable
scaling of $(1-\eta^{2})n^{-1}$. Surprisingly this scaling is
independent of the dimension of the channel.

We have then investigated the multi-use optimal error probability in the
adaptive discrimination of two iso-dimensional Holevo-Werner channels. By
using their teleportation-covariance and the methodology introduced in
Ref.~\cite{Metro}, we have lower- and upper-bounded this optimal probability
by means of single-letter quantities which can be analytically computed from
the associated Werner states. In particular, we have given an explicit formula
for the quantum Chernoff bound, with a similar counterpart for the case of
depolarizing channels.

\bigskip

\textbf{Acknowledgements.~}This work has been supported by the EPSRC via the
`UK Quantum Communications Hub' (EP/M013472/1). T.P.W.C. also acknowledges
funding from a White Rose Scholarship.

\appendix

\section{Quantum Chernoff Bound for Werner states\label{QCB_Werner}}

Let us compute the QCB\ between two arbitrary $d$-dimensional Werner states
$W_{\eta,d}$ and $W_{\zeta,d}$ with $\eta,\zeta\in\lbrack-1,1]$. By definition%
\begin{equation}
Q=\inf_{s\in\lbrack0,1]}Q_{s},~~Q_{s}:=\mathrm{Tr}(W_{\eta,d}^{s}W_{\zeta
,d}^{1-s}).
\end{equation}
Note that we always have $Q_{0}=Q_{1}=1$ so that we may restrict
the infimum in the open interval $s\in(0,1)$. Then, because Werner
states are simultaneously diagonalisable, we may reduce the
computation to
\begin{equation}
Q_{s}=\sum_{i}p_{i}^{s}q_{i}^{1-s}%
\end{equation}
with $p_{i},q_{i}$ being the eigenvalues of $W_{\eta,d}$, and $W_{\zeta,d}$,
respectively. After simple algebra, we obtain
\begin{equation}
Q_{s}=\left(  \frac{1+\zeta}{2}\right)  \left(  \frac{1+\eta}{1+\zeta}\right)
^{s}+\left(  \frac{1-\zeta}{2}\right)  \left(  \frac{1-\eta}{1-\zeta}\right)
^{s}.
\end{equation}

Let us first study singular cases. Firstly, in the scenario where $\eta=\zeta
$, all values of $s$ give identically $Q_{s}=1$, and so we shall define
$s=1/2$ as the optimum for this case. The other cases are:

\begin{itemize}
\item $\zeta=1$; $Q_{s}$ then simplifies to $\left(  \frac{1+\eta}{2}\right)
^{s}$. Since $\frac{1+\eta}{2}\in\lbrack0,1]$, the infimum is achieved for
$s\rightarrow1^{-}$.

\item $\zeta=-1$; $Q_{s}$ then simplifies to $\left(  \frac{1-\eta}{2}\right)
^{s}$. Since $\frac{1-\eta}{2}\in\lbrack0,1]$ as well, the infimum is again
for $s\rightarrow1^{-}$.

\item $\eta=1$; Here $Q_{s}$ simplifies to $\left(  \frac{1+\zeta}{2}\right)
^{1-s}$, thus implying the infimum is achieved for $s\rightarrow0^{+}$.

\item $\eta=-1$; Here $Q_{s}$ simplifies to $\left(  \frac{1-\zeta}{2}\right)
^{1-s}$, and again the infimum is achieved for $s\rightarrow0^{+}$.
\end{itemize}

Once we have studied the previous singular cases (for which the infimum is
taken at the border), let us find the minimum of $Q_{s}$ in the open interval,
where the function is continuous. For simplicity, we will define%
\begin{equation}
k_{\pm}:=\frac{1\pm\zeta}{2},~P:=\frac{1+\eta}{1+\zeta},~M:=\frac{1-\eta
}{1-\zeta}. \label{mlml}%
\end{equation}
so that$\ $%
\begin{equation}
Q_{s}=k_{+}P^{s}+k_{-}M^{s}=k_{+}e^{s\;\mathrm{ln}P}+k_{-}e^{s\;\mathrm{ln}M}.
\end{equation}
Let us now compute the derivative in $s$%
\begin{equation}
\frac{dQ_{s}}{ds}=k_{+}\mathrm{ln}Pe^{s\;\mathrm{ln}P}+k_{-}\mathrm{ln}%
Me^{s\;\mathrm{ln}M}. \label{returnform}%
\end{equation}
By setting $dQ_{s}/ds=0$, we derive
\begin{align}
0  &  =k_{+}\mathrm{ln}Pe^{s\;\mathrm{ln}P}+k_{-}\mathrm{ln}Me^{s\;\mathrm{ln}%
M}\\
k_{+}\mathrm{ln}Pe^{s\;\mathrm{ln}P}  &  =-k_{-}\mathrm{ln}Me^{s\;\mathrm{ln}%
M}\\
\frac{e^{s\mathrm{ln}P}}{e^{s\mathrm{ln}M}}  &  =\frac{-k_{-}\mathrm{ln}%
M}{k_{+}\mathrm{ln}P}\\
e^{s\left(  \mathrm{ln}P-\mathrm{ln}M\right)  }  &  =\frac{-k_{-}\mathrm{ln}%
M}{k_{+}\mathrm{ln}P}\\
s\left(  \mathrm{ln}P-\mathrm{ln}M\right)   &  =\mathrm{ln}\left(
\frac{-k_{-}\mathrm{ln}M}{k_{+}\mathrm{ln}P}\right) \\
s  &  =\frac{\mathrm{ln}\left(  \frac{-k_{-}\mathrm{ln}M}{k_{+}\mathrm{ln}%
P}\right)  }{\mathrm{ln}\left(  \frac{P}{M}\right)  }.
\end{align}
Substituting our definitions in Eq.~(\ref{mlml}), we obtain
\begin{equation}
s=\frac{\mathrm{ln}\left(  \frac{\zeta-1}{\zeta+1}\frac{\mathrm{ln}%
\frac{1-\eta}{1-\zeta}}{\mathrm{ln}\frac{1+\eta}{1+\zeta}}\right)
}{\mathrm{ln}\frac{\left(  1+\eta\right)  \left(  1-\zeta\right)  }{\left(
1+\zeta\right)  \left(  1-\eta\right)  }}=:s_{\eta,\zeta}. \label{sval}%
\end{equation}

It remains to be proven that the critical point $s_{\eta,\zeta}$ is in
$[0,1]$. First we shall prove that $s_{\eta,\zeta}$ is positive. We shall
start by considering the denominator, in two scenarios:

\begin{itemize}
\item $-1<\zeta<\eta<1$. In this scenario, both fractions $\frac{1+\eta
}{1+\zeta}$ and $\frac{1-\zeta}{1-\eta}$ must necessarily be
greater than 1; thus the overall denominator is the logarithm of
something greater than 1, and therefore positive.

\item $-1<\eta<\zeta<1$. Conversely, in this case both fractions are
$\frac{1+\eta}{1+\zeta}$ and $\frac{1-\zeta}{1-\eta}$ are less
than one, but positive, and so too is their product; forcing the
overall denominator to be negative when the logarithm is taken.
\end{itemize}

In order for $s$ to be positive in all scenarios, this means we
require:

\begin{itemize}
\item For $-1<\zeta<\eta<1$, the numerator is positive; equivalently we
require that:
\begin{equation}
\frac{\zeta-1}{\zeta+1}\frac{\mathrm{ln}\frac{1-\eta}{1-\zeta}}{\mathrm{ln}%
\frac{1+\eta}{1+\zeta}}\geq1. \label{firstcaseapp}%
\end{equation}
Since $\zeta+1>0$, and $\frac{1+\eta}{1+\zeta}>1$, we have the denominator of
Eq.~(\ref{firstcaseapp}) is positive, so the equation can be rearranged to
give:
\begin{equation}
\left(  \zeta-1\right)  \mathrm{ln}\frac{1-\eta}{1-\zeta}-\left(
\zeta+1\right)  \mathrm{ln}\frac{1+\eta}{1+\zeta}\geq0.
\end{equation}

\item Similarly we require the numerator to be negative if $-1<\eta<\zeta<1$,
which is equivalent to
\begin{equation}
\frac{\zeta-1}{\zeta+1}\frac{\mathrm{ln}\frac{1-\eta}{1-\zeta}}{\mathrm{ln}%
\frac{1+\eta}{1+\zeta}}\leq1.\label{secondcaseapp}%
\end{equation}
This time, although $\zeta+1$ is still positive, we have that $\frac{1+\eta
}{1+\zeta}<1$, and therefore the denominator of Eq.~(\ref{secondcaseapp}) is
negative. This means, when we multiply out the denominator of
(\ref{secondcaseapp}) we obtain
\begin{equation}
\left(  \zeta-1\right)  \mathrm{ln}\frac{1-\eta}{1-\zeta}-\left(
\zeta+1\right)  \mathrm{ln}\frac{1+\eta}{1+\zeta}\geq0.\label{endpoint}%
\end{equation}

\end{itemize}

We see that, regardless which of $\eta,\zeta$ is greater, we require the same
statement. First note that Eq.~(\ref{endpoint}) is true if and only if
\begin{equation}
\left(  \frac{1-\zeta}{2}\right)  \mathrm{ln}\left(  \frac{\frac{1-\zeta}{2}%
}{\frac{1-\eta}{2}}\right)  +\left(  \frac{\zeta+1}{2}\right)  \mathrm{ln}%
\left(  \frac{\frac{1+\zeta}{2}}{\frac{1+\eta}{2}}\right)  \geq0.
\end{equation}
We then make a substitution of variables $p_{\eta}=\frac{1+\eta}{2}$ and
$p_{\zeta}=\frac{1+\zeta}{2}$, so that left hand side becomes
\begin{equation}
(1-p_{\zeta})\mathrm{ln}\frac{1-p_{\zeta}}{1-p_{\eta}}+p_{\zeta}%
\mathrm{ln}\frac{p_{\zeta}}{p_{\eta}}%
\end{equation}
and $p_{\eta},p_{\zeta}\in(0,1)$. This is simply the classical relative
entropy, or Kullback-Leibler (KL) divergence, in a different logarithmic base,
of two biased coin flips. However, Gibbs inequality states that, for any
logarithmic basis, the KL-divergence is always non-negative. Thus we have
necessarily that $s_{\eta,\zeta}$ is non-negative for all value of $\eta
,\zeta$. To show that $s_{\eta,\zeta}\leq1$, we shall instead prove a stronger
result, that $s_{\eta,\zeta}+s_{\zeta,\eta}=1$. Since both terms are
non-negative, this is a sufficient statement.

Using our formula in Eq.~(\ref{sval}), we may write
\begin{equation}
s_{\eta,\zeta}+s_{\zeta,\eta}=\frac{\mathrm{ln}\left(  \frac{\zeta-1}{\zeta
+1}\frac{\mathrm{ln}\frac{1-\eta}{1-\zeta}}{\mathrm{ln}\frac{1+\eta}{1+\zeta}%
}\right)  }{\mathrm{ln}\frac{\left(  1+\eta\right)  \left(  1-\zeta\right)
}{\left(  1+\zeta\right)  \left(  1-\eta\right)  }}+\frac{\mathrm{ln}\left(
\frac{\eta+1}{\eta-1}\frac{\mathrm{ln}\frac{1+\zeta}{1+\eta}}{\mathrm{ln}%
\frac{1-\zeta}{1-\eta}}\right)  }{\mathrm{ln}\frac{\left(  1+\eta\right)
\left(  1-\zeta\right)  }{\left(  1+\zeta\right)  \left(  1-\eta\right)  }}.
\end{equation}
Since they share a denominator, we shall look at the numerator, which can be
simplified as follows
\begin{align}
&  \mathrm{ln}\left(  \frac{\zeta-1}{\zeta+1}\frac{\mathrm{ln}\frac{1-\eta
}{1-\zeta}}{\mathrm{ln}\frac{1+\eta}{1+\zeta}}\right)  +\mathrm{ln}\left(
\frac{\eta+1}{\eta-1}\frac{\mathrm{ln}\frac{1+\zeta}{1+\eta}}{\mathrm{ln}%
\frac{1-\zeta}{1-\eta}}\right) \nonumber\\
&  =\mathrm{ln}\frac{\left(  1+\eta\right)  \left(  1-\zeta\right)  }{\left(
1+\zeta\right)  \left(  1-\eta\right)  },
\end{align}
where we have used $1=-1^{2}$, and absorbed the minus signs into either the
brackets or logarithms.

Thus we must conclude that $s_{\eta,\zeta}+s_{\zeta,\eta}=1$, and with that,
we may conclude that for all valid values of $\eta,\zeta$, our given
$s_{\eta,\zeta}$ is within the region $[0,1]$. Moreover, it satisfied
$\frac{dQ_{s}}{ds}|_{s=s_{\eta,\zeta}}=0$. We need only that $\frac{d^{2}%
Q_{s}}{ds^{2}}|_{s=s_{\eta,\zeta}}>0$, to show it is a minima. To do this, we
return to Eq.~(\ref{returnform}). Differentiating again, we see that
\begin{equation}
\frac{d^{2}Q_{s}}{ds^{2}}=k_{+}[\mathrm{ln}(P)]^{2}e^{s\mathrm{ln}(P)}%
+k_{-}[\mathrm{ln}(M)]^{2}e^{s\mathrm{ln}(M)}.
\end{equation}
When $\eta,\zeta\in(0,1)$, we have that $k_{+},k_{1}$ are strictly
positive, and when $\eta\neq\zeta$ that
$\mathrm{ln}(P),\mathrm{ln}(M)\neq0$ and thus their squares are
positive too. Finally, $e$ to the power of any real value is
strictly positive, and thus we have $\frac{d^{2}Q_{s}}{ds^{2}}>0$
for all values of $s$, including $s_{\eta,\zeta}$. Thus we have
proven, in combination with the special cases stated above, that
our stated $s$ truly minimises the QCB.

\section{Quantum Chernoff Bound for Isotropic states\label{QCB_iso}}

Let us compute the QCB\ between two arbitrary $d$-dimensional isotropic states
$\Omega_{\alpha,d}$ and $\Omega_{\beta,d}$ with $\alpha,\beta\in\lbrack0,d]$.
By restricting definition of QCB to the open interval $s\in(0,1)$, we write%
\begin{equation}
Q=\inf_{s\in(0,1)}Q_{s},~~Q_{s}:=\mathrm{Tr}(\Omega_{\alpha,d}^{s}%
,\Omega_{\beta,d}^{1-s}).
\end{equation}
As a consequence of the isotropic states being simultaneously
diagonalisable, we may rewrite $Q_{s}$ as
\begin{equation}
Q_{s}=\frac{\beta}{d}\left(  \frac{\alpha}{\beta}\right)  ^{s}+\frac{d-\beta
}{d}\left(  \frac{d-\alpha}{d-\beta}\right)  ^{s}.
\end{equation}

First of all, let us study the singular cases. We have:

\begin{itemize}
\item $\beta=d\Rightarrow$ infimum at $s\rightarrow1^{-}$.

\item $\beta=0\Rightarrow$ infimum at $s\rightarrow1^{-}$.

\item $\alpha=d\Rightarrow$ infimum at $s\rightarrow0^{+}$.

\item $\alpha=0\Rightarrow$ infimum at $s\rightarrow0^{+}$.

\item $\alpha=\beta$ $\Rightarrow$ minimum at $s=1/2$.
\end{itemize}

We then compute the derivative, which is given by%
\begin{equation}
\frac{dQ_{s}}{ds}=l_{+}\mathrm{ln}P_{\Omega}e^{s\mathrm{ln}P_{\Omega}}%
+l_{-}\mathrm{ln}M_{\Omega}e^{s\mathrm{ln}M_{\Omega}}.
\end{equation}
where we have set%
\begin{equation}
l_{+}=\frac{\beta}{d},~l_{-}=\frac{d-\beta}{d},~P_{\Omega}=\frac{\alpha}%
{\beta},~M_{\Omega}=\frac{d-\alpha}{d-\beta}.
\end{equation}
From $dQ_{s}/ds=0$, we compute the critical point, obtaining
\begin{equation}
s=\frac{\mathrm{ln}\left(  \frac{\beta-d}{\beta}\frac{\mathrm{ln}%
\frac{d-\alpha}{d-\beta}}{\mathrm{ln}\frac{\alpha}{\beta}}\right)
}{\mathrm{ln}\frac{\alpha(d-\beta)}{\beta(d-\alpha)}}=:s_{\alpha,\beta
}^{\Omega}.
\end{equation}
Unfortunately $s_{\alpha,\beta}^{\Omega}$ is dimension-dependent. We may
transform $\alpha,\beta$ to dimension independent variables by setting
$\eta=\frac{2\alpha-d}{d}\in\lbrack-1,1]$ and $\zeta=\frac{2\beta-d}{d}%
\in\lbrack-1,1]$. When these are substituted into $s_{\alpha,\beta}^{\Omega}$,
we find $s_{\alpha,\beta}^{\Omega}=s_{\eta,\zeta}$, where $s_{\eta,\zeta}$ is
the one defined in Eq.~(\ref{sval}) that we already know to be a minimum in
the open interval.


\begin{thebibliography}{99}                                                                                               %


\bibitem {NiCh}M. A. Nielsen, and I. L. Chuang, \textit{Quantum Computation
and Quantum Information} (Cambridge University Press, Cambridge, 2000).

\bibitem {QIbook}M. Hayashi, \textit{Quantum Information Theory: Mathematical
Foundation} (Springer-Verlag Berlin Heidelberg, 2017).

\bibitem {RMP}C. Weedbrook, S. Pirandola, R. Garc\'{\i}a-Patr\'{o}n, N. J.
Cerf, T. C. Ralph, J. H. Shapiro, and S. Lloyd, Rev. Mod. Phys. \textbf{84},
621 (2012).

\bibitem {BraRMP}S. L. Braunstein and P. van Loock, Rev. Mod. Phys.
\textbf{77}, 513 (2005).

\bibitem {Ulrikreview}U. L. Andersen, J. S. Neergaard-Nielsen, P. van Loock,
and A. Furusawa, Nature Phys. \textbf{11}, 713 (2015).

\bibitem {Kimble}H. J. Kimble, Nature \textbf{453}, 1023 (2008).

\bibitem {HybridINTERNET}S. Pirandola, and S. L. Braunstein, Nature
\textbf{532}, 169 (2016).

\bibitem {crypt1}C. H. Bennett and G. Brassard. \textit{Quantum Cryptography:
Public Key Distribution and Coin Tossing}, Proceedings of IEEE International
Conference on Computers, Systems and Signal Processing, \textbf{175} (1984).

\bibitem {crypt2}A.K. Ekert, Phys. Rev. Lett. \textbf{67}, 661 (1991).

\bibitem {crypt3}N. Gisin, G. Ribordy, W. Tittel, and H. Zbinden, Rev. Mod.
Phys. \textbf{74}, 145 (2002).

\bibitem {Grosshans2003b}F. Grosshans, G. Van Ache, J. Wenger, R. Brouri, N.
J. Cerf, and P. Grangier, Nature \textbf{421}, 238 (2003).

\bibitem {weedbrook2004noswitching}C. Weedbrook, A. M. Lance, W. P. Bowen, T.
Symul, T. C. Ralph, and P. K. Lam, Phys. Rev. Lett. \textbf{93}, 170504 (2004).

\bibitem {crypt5}R. Colbeck, \textit{Quantum And Relativistic Protocols For
Secure Multi-Party Computation} (PhD thesis, University of Cambridge, 2006).

\bibitem {pirs2way}S. Pirandola, S. Mancini, S. Lloyd, and S. L. Braunstein,
Nat. Phys. \textbf{4}, 726 (2008).

\bibitem {MDI1}S. L. Braunstein and S. Pirandola, Phys. Rev. Lett.
\textbf{108}, 130502 (2012).

\bibitem {MDILo}M.Curty, B. Qi, H.K. Lo, Phys. Rev. Lett. \textbf{108}, 130503 (2012).

\bibitem {CVMDIQKD}S. Pirandola, C. Ottaviani, G. Spedalieri, C. Weedbrook, S.
L. Braunstein, S. Lloyd, T. Ghering, C.S. Jacobsen, and U. L.
Andersen\textit{,} Nat. Photon. \textbf{9}, 397 (2015).

\bibitem {diamantiENTROPY}E. Diamanti and A. Leverrier, Entropy \textbf{17},
6072 (2015).

\bibitem {usenkoREVIEW}V. C. Usenko and R. Filip, Entropy \textbf{18}, 20 (2016).

\bibitem {comp1}P. Shor, \textit{Polynomial-Time Algorithms for Prime
Factorization and Discrete Logarithms on a Quantum Computer}, Proceedings of
the 35th Annual Symposium on Foundations of Computer Science, Santa Fe (1994).

\bibitem {comp4}S. Lloyd and S. L. Braunstein, Phys. Rev. Lett. \textbf{82},
1784 (1999).

\bibitem {comp5}T. D. Ladd, F. Jelezko, R. Laflamme,Y. Nakamura, C. Monroe and
J. L. O'Brien, Nature \textbf{464}, 45 (2010).

\bibitem {met1}S. L. Braunstein and C. M. Caves, Phys. Rev. Lett. \textbf{72},
3439 (1994).

\bibitem {met2}P. Kok, S. L. Braunstein and J. P. Dowling, J. Op. B
\textbf{6}, 8 (2004).

\bibitem {met3}V. Giovannetti, S. Lloyd and L. Maccone, Science \textbf{306},
1330 (2004).

\bibitem {met4}H. M. Wiseman and G. J. Milburn, \textit{Quantum Measurement
and Control} (Cambridge University Press, 2010).

\bibitem {met6}V. Giovannetti, S. Lloyd, and L. Maccone, Nature Photonics
\textbf{5}, 222--229 (2011).

\bibitem {met5}G. Toth, I. Apellaniz, J. Phys. A: Math. Theor. \textbf{47},
424006 (2014).

\bibitem {Paris}M. G. A. Paris, Int. J. Quant. Inf. \textbf{7}, 125 (2009).

\bibitem {Makei}M. Tsang, R. Nair, and X. Lu, Phys. Rev. X \textbf{6}, 031033 (2016).

\bibitem {Lupo16}C. Lupo and S. Pirandola, Phys. Rev. Lett. \textbf{117},
190802 (2016).

\bibitem {Nair}R. Nair, and M. Tsang, Phys. Rev. Lett. \textbf{117}, 190801 (2016).

\bibitem {Metro}S. Pirandola, and C. Lupo, Phys. Rev. Lett. \textbf{118},
100502 (2017); \textit{ibid.} \textbf{119}, 129901 (2017).

\bibitem {QHT}A. Chefles, Contemp. Phys. \textbf{41}, 401 (2000).

\bibitem {QHT2}S. M. Barnett and S. Croke, Advances in Optics and Photonics
\textbf{1}, 238 (2009).

\bibitem {Invernizzi}C. Invernizzi, M. G. A. Paris, and S. Pirandola, Phys.
Rev. A \textbf{84}, 022334 (2011).

\bibitem {QHB1}K. M. R. Audenaert, M. Nussbaum, A. Szkola, and F. Verstraete,
Commun. Math. Phys. \textbf{279}, 251 (2008).

\bibitem {Gae1}G. Spedalieri and S. L. Braunstein, Phys. Rev. A \textbf{90},
052307 (2014).

\bibitem {Qread}S. Pirandola, Phys. Rev. Lett. \textbf{106}, 090504 (2011).

\bibitem {Qread2}S.~Pirandola, C.~Lupo, V.~Giovannetti, S.~Mancini, and S.
L.~Braunstein, New J. Phys. \textbf{13}, 113012 (2011).

\bibitem {Qread3}G. Spedalieri, C. Lupo, S. Mancini, S. L. Braunstein, and S.
Pirandola, Phys. Rev. A \textbf{86}, 012315 (2012).

\bibitem {Qread4}C. Lupo, S. Pirandola, V. Giovannetti, and S. Mancini, Phys.
Rev. A \textbf{87}, 062310 (2013).

\bibitem {Nair11}R.~Nair, Phys. Rev. A \textbf{84}, 032312 (2011).

\bibitem {Hirota11}O.~Hirota, arXiv:1108.4163 (2011).

\bibitem {Bisio11}A.~Bisio, M.~Dall'Arno, and G. M.~D'Ariano, Phys. Rev. A
\textbf{84}, 012310 (2011).

\bibitem {Arno11}M.~Dall'Arno \textit{et al.}, Phys. Rev. A \textbf{85},
012308 (2012).

\bibitem {Qill0}S. Lloyd, Science \textbf{321}, 1463 (2008).

\bibitem {Qill1}S.-H. Tan \textit{et al.}, Phys. Rev. Lett. \textbf{101},
253601 (2008).

\bibitem {Qill2}S. Barzanjeh \textit{et al.}, Phys. Rev. Lett. \textbf{114},
080503 (2015).

\bibitem {Qill3}C. Weedbrook, S. Pirandola, J. Thompson, V. Vedral, and M. Gu,
New J. Phys. \textbf{18}, 043027 (2016).

\bibitem {Qill4}E. D. Lopaeva, I. Ruo Berchera, I. P. Degiovanni, S. Olivares,
G. Brida, and M. Genovese, Phys. Rev. Lett. \textbf{110}, 153603 (2013).

\bibitem {Qill5}Z. Zhang, S. Mouradian, F. N.C. Wong, and J. H. Shapiro, Phys.
Rev. Lett. \textbf{114}, 110506 (2015).

\bibitem {Hesltrom}C. W. Helstrom, \textit{Quantum Detection and Estimation
Theory} (New York: Academic, 1976).

\bibitem {Fid1}A. Uhlmann, Rep. Math. Phys. \textbf{9}, 273 (1976).

\bibitem {Fid2}R. Jozsa, Journal of Modern Optics \textbf{41}, 2315 (1994).

\bibitem {Banchi}L. Banchi, S. L. Braunstein, and S. Pirandola, Phys. Rev.
Lett. \textbf{115}, 260501 (2015).

\bibitem {QCB1}K. M. R. Audenaert\textit{ et al.}, Phys. Rev. Lett.
\textbf{98}, 160501 (2007).

\bibitem {QCB2}J. Calsamiglia, R. Munoz-Tapia, L. Masanes, A. Acin, and E.
Bagan, Phys. Rev. A \textbf{77}, 032311 (2008).

\bibitem {QCB3}S. Pirandola, and S. Lloyd, Phys. Rev. A \textbf{78}, 012331 (2008).

\bibitem {tele}C. H. Bennett, G. Brassard, C. Crepeau, R. Jozsa, A. Peres, and
W. K. Wootters, Phys. Rev. Lett. \textbf{70}, 1895 (1993).

\bibitem {Samtele}S. L. Braunstein, and H. J. Kimble, Phys. Rev. Lett.
\textbf{80}, 869 (1998).

\bibitem {teleREVIEW}S. Pirandola, J. Eisert, C. Weedbrook, A. Furusawa, and
S. L. Braunstein, Nat. Photon. \textbf{9}, 641 (2015).

\bibitem {Stretching}S. Pirandola, R. Laurenza, C. Ottaviani and L. Banchi,
Nat. Comm. \textbf{8}, 15043 (2017). See also arXiv:1510.08863 (2015).

\bibitem {RevewTQC}S. Pirandola, S. L. Braunstein, R. Laurenza, C. Ottaviani,
T. P. W. Cope, G. Spedalieri, and L. Banchi, \textit{Theory of Channel
Simulation and Bounds for Private Communication}, arXiv:1711.09909v1 (2017).

\bibitem {Werner}R. F. Werner, Phys. Rev. A, \textbf{40}, 4277 (Oct 1989).

\bibitem {Barrett}F. Barrett, Phys. Rev. A \textbf{65}, 042302 (2002).

\bibitem {NPTstates}D. Z. Djokovic, Entropy \textbf{18}, 216 (2016).

\bibitem {counter}R. F. Werner, and A.S. Holevo, J. Mat. Phys. \textbf{43},
4353 (2002).

\bibitem {CRbound}S. L. Braustein, and C. M. Caves, G. J. Milburn, Ann. Phys.
\textbf{247}, 135 (1996).

\bibitem {VolWer}K. G. H. Vollbrecht and R. F. Werner, Phys. Rev. A
\textbf{64}, 062307 (2001).

\bibitem {VedFORM}V. Vedral, M. B. Plenio, M. A. Rippin, and P. L. Knight,
Phys. Rev. Lett. \textbf{78}, 2275 (1997).

\bibitem {Plenio}V. Vedral, and M. B. Plenio, Phys. Rev. A \textbf{57}, 1619 (1998).

\bibitem {VedralRMP}V. Vedral, Rev. Mod. Phys. \textbf{74}, 197 (2002).

\bibitem {Fanneschannel}M. Fannes, B. Haegeman,M. Mosonyi and D. Vanpeteghem,
arXiv:quant-ph/0410195 (2004).

\bibitem {horoISO}M. Horodecki and P. Horodecki, Phys. Rev. A \textbf{59},
4206 (1999).

\bibitem {MHthesis}A. Muller-Hermes, \textit{Transposition in Quantum
Information Theory} (Master's thesis, Technical University of Munich, 2012).

\bibitem {Wolfnotes}M. M. Wolf, Notes on \textquotedblleft Quantum Channels \&
Operations\textquotedblright\ (see page 35). Available at
https://www-m5.ma.tum.de/foswiki/pub/M5/Allgemeines/ MichaelWolf/QChannelLecture.pdf.

\bibitem {Leung}D. Leung and W. Matthews, IEEE Trans. Info. Theory
\textbf{61}, 4486 (2015).

\bibitem {ref1}S. Pirandola, \textit{Capacities of Repeater-Assisted Quantum
Communications}, arXiv:1601.00966 (2016).

\bibitem {multipoint}R. Laurenza and S. Pirandola, Phys. Rev. A \textbf{96},
032318 (2017).

\bibitem {RicFINITE}R. Laurenza, S. L. Braunstein, and S. Pirandola,
\textit{Finite-Resource Teleportation Stretching for Continuous-Variable
Systems}, arXiv:1706.06065 (2017).

\bibitem {Tompaper}T. P. W. Cope, L. Hetzel, L. Banchi, and S. Pirandola,
Phys. Rev. A \textbf{96}, 022323 (2017).

\bibitem {ReviewMETRO}R. Laurenza, C. Lupo, G. Spedalieri, S. L. Braunstein,
and S. Pirandola, \textit{Channel Simulation in Quantum
Metrology}, arXiv:1712.06603 (2017).

\bibitem {Gillapp}R. D. Gill and S. Massar, Phys. Rev. A \textbf{61}, 042312 (2000).
\end{thebibliography}
\end{document}